\documentclass[pre,amsmath,amssymb,twocolumn,showpacs]{revtex4-1}
\usepackage{bm}
\usepackage{amssymb}
\usepackage{colordvi}
\usepackage{graphicx}
\usepackage{color}
\usepackage{hyperref}
\newcommand{\up}{\uparrow}
\newcommand{\down}{\downarrow}

\newcommand{\be}{\begin{equation}}
\newcommand{\ee}{\end{equation}}
\newcommand{\bea}{\begin{eqnarray}}
\newcommand{\eea}{\end{eqnarray}}
\newcommand{\ep}{\epsilon}

\newcommand{\ave}[1]{\langle #1\rangle}
\newcommand{\ome}{\omega}
\newcommand{\Ome}{\Omega}

\def\nn{\nonumber}

\def\grad {\mbox{\boldmath$\nabla$\unboldmath}}

\begin{document}

\title{Thermodynamic bounds and general properties of optimal 
  efficiency and power in linear responses}

\author{Jian-Hua Jiang}\email{jianhua.jiang.phys@gmail.com}
\affiliation{Department of Physics, University of Toronto, Toronto,
  Ontario, M5S 1A7 Canada}

\date{\today}

\begin{abstract}
We study the optimal exergy efficiency and power for  thermodynamic
systems with Onsager-type ``current-force'' relationship describing
the linear-response to external influences. We derive, in 
analytic forms, the maximum efficiency and optimal efficiency for
maximum power for a thermodynamic machine 
described by a $N\times N$ symmetric Onsager matrix with arbitrary
$N$. The figure of merit is expressed in
terms of the largest eigenvalue of the ``coupling matrix'' which is
solely determined by the Onsager matrix. Some simple but general relationships
between the power and efficiency at the conditions for (i) maximum
efficiency and (ii) optimal efficiency for maximum power are obtained.
We show how the second law of thermodynamics bounds the optimal
efficiency and the Onsager matrix, and relate those bounds together.
{The maximum power theorem (Jacobi's Law) is generalized to all
thermodynamic machines with symmetric Onsager matrix in the linear-response regime.}
We also discuss systems with asymmetric Onsager matrix (such as
systems under magnetic field) {for a particular situation and we
  show that} the
reversible limit of efficiency can be reached at finite output
power. Cooperative effects are found 
to improve the figure of merit significantly in systems with 
multiply cross-correlated responses. Application to example systems
demonstrates that the theory is helpful in guiding the search for high
performance materials and structures in energy researches.
\end{abstract}

% Only one pacs!!!
% 05.70.Ln : irreversible thermodynamics

% --------------- emphasize novelty and universality

\maketitle

\section{Introduction}
Under challenges imposed by increasing demand yet limited availability
of energy resources, improving energy efficiency becomes increasingly
important in technology developments. Historically, Carnot deduced 
that for a heat engine operating between two reservoirs with
temperatures $T_h$ and $T_c$ ($T_h>T_c$), the energy conversion
efficiency, $\eta=W/Q$ ($W$ is the output work, $Q$ is the heat from
the hot reservoir), has a maximum value, namely the Carnot efficiency,
$\eta_C=(T_h-T_c)/T_h$\cite{callen}. The Carnot efficiency is only for
ideal machines operating in the reversible limit. Energy efficiency of
realistic machines is reduced by unavoidable irreversible
entropy production. A way to count the reduction of energy efficiency
from the value at reversible limit is to use the exergy efficiency (or
``second-law efficiency'')\cite{odum,2ndlaw1,2ndlaw2,2ndlawbook,caplan1,caplan2} 
\be
\phi \equiv \frac{\dot {\cal A}_{out}}{\dot {\cal A}_{in}} \label{phi}
\ee
where $\dot {\cal A}_{out}$ and $\dot {\cal A}_{in}$ are the output
and input {\em exergy} (i.e., the Gibbs free energy) per unit
time. Exergy is defined as ${\cal A}={\cal U}-TS$ where ${\cal U}$ is
the enthalpy (i.e., the total energy), $T$ is the temperature, and $S$
is the entropy. Although the total energy is conserved, the output exergy is
reduced by entropy production, $\dot S_{tot}$, as $\dot {\cal A}_{out}=\dot
{\cal A}_{in}-T\dot S_{tot}$, hence $\phi\le 1$. Both $\phi\le 1$ and
$\eta\le \eta_C$ are dictated by the second law of thermodynamics. In
fact for thermoelectric engine and refrigerator the two are related
by\cite{odum,2ndlawbook,caplan1,ca-ex,Seifert-review} 
\be
\phi=\frac{\eta}{\eta_C} .
\ee
For this reason, exergy efficiency is also called as
``rational efficiency''\cite{2ndlawbook}. Using Onsager's theory of 
irreversible thermodynamics and the exergy efficiency, the study of
efficiency of heat engines, chemical engines, and other energy devices
can be presented in an uniform
manner\cite{ca-ex,Seifert-review,odum,2ndlaw1,2ndlaw2,2ndlawbook,caplan1,caplan2}. Specifically,
the efficiency of chemical engines, the output work divided by the chemical work, is
precisely Eq.~(\ref{phi}), as the output work is equal to the output
exergy and the input chemical work is equal to the input (consumed)
exergy\cite{ca-ex,Seifert-review,2ndlaw1}. The exergy efficiency becomes particularly
convenient for machines with multiple forms of input (or output)
energy\cite{caplan2}. For example in a spin-thermoelectric\cite{bauer}
refrigerator, both electrical energy and magnetic energy are consumed
to drive the cooling (see Sec.~\ref{sec:s-th-cooling}).

A central issue in energy application is to find out the optimal
efficiency and maximum power of a machine and the conditions that
realize them\cite{devos,ca,nicolis}. For example, Ioffe derived the optimal exergy efficiency for
isotropic thermoelectric materials in the linear-response regime
as\cite{ioffe}
\be
\eta_{max} = \eta_C \frac{\sqrt{\xi+1}-1}{\sqrt{\xi+1}+1}, \quad\quad \xi = \frac{\sigma
S^2 T}{\kappa} .\label{zt}
\ee
The figure of merit, $\xi$, is {\em solely} determined by the transport
coefficients of the material: the electrical conductivity $\sigma$,
the Seebeck coefficient $S$, and the thermal conductivity
$\kappa$. This property is an important guiding
principle in the search of high performance thermoelectric
materials\cite{honig,ms,review}.

However, Eq.~(\ref{zt}) was derived for isotropic systems, where, by
choosing a proper set of coordinate axes, the problem can be reduced
to correlated transport for two scalar currents: one heat current and
one electric current. Quite often in anisotropic materials, the
complete description of thermoelectric transport must involve six
scalar currents as both the electric and heat currents consist of
three scalar components (e.g., the electrical current
$\vec{j}=(j_x,j_y,j_z)$ with $j_x$, $j_y$, and $j_z$ being the 
components in the $x$, $y$, and $z$ directions)\cite{honig,aniso}. For
piezoelectric energy conversion in an anisotropic material, the full
description of responses involves nine scalar ``currents'': three of them
are electric displacements and the other six are strains\cite{pe}.
The description of these cross-correlated responses can be simplified
only for certain high symmetry structures. Recent development of
technologies for high-quality thin film growth which allows precise
control of composition, atomic arrangements and interfaces provides
the toolbox for functional nano-structured composite materials which
could have pronounced application values that does not share by their
compounds. Often these composite structures have lower symmetry and
the full description of cross-correlated responses cannot be
simplified. Besides, breaking time-reversal symmetry brings
further complication to cross-correlated responses\cite{trb1,trb2,trb3}.
Quite often Ioffe's derivation of optimal energy efficiency cannot be
directly applied to those practical systems. In those situations the
(global) maximum efficiency is rather difficult to find, although one
can always easily find certain optimal efficiencies under
restrictions\cite{honig,aniso,multi,pe,caplan2}.

Finding the optimal exergy efficiency and power for complex thermodynamic
systems has stimulated a number of studies\cite{aniso,bio,caplan2}. It
becomes increasingly important as researches reveal more
cross-correlated responses and realize their
applications\cite{bauer,zlwang1,zlwang2}. Fast developing
nanotechnologies and 
material technologies offer a large number of materials and structures
of which complex cross-correlated responses are enhanced and made
available for practical applications. Examples are, spin-thermoelectric
effect\cite{bauer}, piezopotential gating\cite{zlwang1} and
piezo-phototronics\cite{zlwang2}, to name but just a few.
Besides, biological systems are often characterized by
cross-correlated responses to density, temperature, and electrochemical
potential gradients\cite{bio,caplan2}. A typical example is transport
across a biological membrane: even for a single ionic solution,
transmission through the membrane must be described by three flows,
the volume flow, the solute flow, and the electrical flow, which are
often cross-correlated\cite{bio}. Cross-correlated responses enable
energy conversion from one form to another, during which the functions
of a machine is realized (a ``machine'' is a system which consumes
input energy to achieve a practical goal by doing work to the
external). Caplan derived the analytic expression of the optimal
exergy efficiency for machines with only one flow for 
energy input but multiple flows for output or vice
versa\cite{caplan2}. However, general results on the optimal
efficiency and power are still absent, particularly in analytic forms.

In this work we derive analytic results for optimal efficiency and
power under general considerations that can be applied to a broad
range of thermodynamic systems. The requirements are only that there
exists an Onsager-type ``current-force'' relation that describes the
responses to external influences (``forces'')\cite{onsager} and that
the system is operating at  steady states in linear responses.
These requirements are often satisfied for physical systems with 
forces not too strong\cite{groot,nicolis}. The derived results can be
connected with realistic systems of which the output power is
consumed by a device or by large a power grid. We obtain some 
simple but general relationships that connect the optimal power and
efficiency for different optimization schemes. These results are first
obtained for systems with symmetric Onsager response matrix and then
extended to systems with asymmetric Onsager matrix (e.g., systems
under magnetic field). We point out that cooperative effects can be used to improve
efficiency (figure of merit) for systems with multiple
cross-correlated responses. Such improvement, achieved via combining
different input (or output) forces rather than engineering materials, can
be significant in systems with multiple cross-correlated responses.
Examples are given to demonstrate how the theory is used to guide the
search for high performance energy applications.

This paper is organized as follows: in Sec.~II we establish the basic
formalism by using Onsager's theory of irreversible thermodynamic
processes in the linear-response regime. We derive the optimal
efficiency and output power for symmetric Onsager matrix in
Sec.~III. In Sec.~IV the derivation is re-interpreted with realistic
considerations where parasitic dissipation and the response of the
device accepting the output energy are considered. We extend the study
to systems with asymmetric Onsager matrix in Sec.~V. Examples 
that illustrate the usefulness of the findings are presented in
Sec.~VI, and we conclude in Sec.~VII.

\section{Basic formalism}
Under external influences (``forces'') a thermodynamic system develops
motions that deviate from their equilibrium values. These motions
(``currents'') can be described quantitatively by the rates of changes
in thermodynamic state variables\cite{groot,landau}. The relation
between the forces $\vec{{\cal F}}$ and currents $\vec{{\cal J}}$ is
generally written as\cite{groot,landau}
\be
\vec{{\cal J}} = \hat{{\cal M}}\vec{{\cal F}}\quad {\rm or}\quad {\cal J}_n= \sum_{k} {\cal M}_{nk} {\cal F}_k ,\label{c-f}
\ee
where the index $n$ ($k$) numerates all currents (forces), and
$\hat{{\cal M}}$ is the Onsager matrix. When the forces are not too
strong the dependence of $\hat{{\cal M}}$ on the forces can be
ignored. Cross-correlated responses (e.g.,
thermoelectric effect) allow energy conversion from the input forms to
the output forms and realize functions of a machine. According to the theory
of irreversible thermodynamics\cite{onsager,groot}, there are an equal
number of forces and currents. Each force ${\cal F}_n$ has a
conjugated current ${\cal J}_n$ such that the reduction of total
exergy (Gibbs free energy) is given by
\be
- \dot {\cal A}_{tot} = T\dot S_{tot} = \sum_n {\cal J}_n {\cal F}_n .
\ee
The reduction of exergy $-\dot {\cal A}_n = {\cal J}_n {\cal F}_n$
associated with the current ${\cal J}_n$ for exergy input 
is positive, while for exergy output it is negative. Hence the input
and output exergy are\cite{caplan2} 
\be
\dot {\cal A}_{in}\equiv \sum_{n\in I} {\cal J}_n {\cal F}_n,\quad \dot {\cal A}_{out}\equiv - \sum_{k\in O} {\cal J}_k {\cal F}_k ,
\ee
respectively. The sets $I$ and $O$ in the above refer to exergy input and
output, respectively. The output exergy is also the output work, i.e.,
$\dot W = \dot {\cal A}_{out}$ (Throughout this paper ``work'' is
associated with linear-response processes for given thermodynamic
forces, i.e., work and efficiency are functions of thermodynamic
forces). For $\dot {\cal A}_{in}>0$ the exergy efficiency is
\be
\phi = \frac{-\sum_{k\in O} {\cal J}_k {\cal F}_k }{\sum_{n\in I}
  {\cal J}_n {\cal F}_n} = \frac{\dot {\cal A}_{in}-T\dot
  S_{tot}}{\dot {\cal A}_{in}} \le 100\% . \label{phi-def}
\ee
Only in the reversible limit, $\dot S_{tot}=0$, the exergy efficiency $\phi$
reaches its upper bound. The second law of thermodynamics requires
$\dot S_{tot}\ge 0$ for all possible values of forces. This is
satisfied only when {\em all} eigenvalues of the Onsager matrix
$\hat{{\cal M}}$ are positive [see Appendix~A] (note that, as the
reversible limit, $\dot S_{tot}= 0$, does not exist for realistic
systems, we consider only situations with positive entropy production.
Zero entropy production is the limit when the entropy
production is extremely small. In this way, {\em all} eigenvalues of
the Onsager matrix must be greater than zero.)  This property is
briefly stated as that {\em Onsager matrix is positive}.

\section{Optimizing efficiency and power for systems with symmetric
  Onsager matrix}
\label{op1}

The maximum exergy efficiency is obtained by solving the differential equation
\be
\partial_{{\cal F}_k} \phi = 0,\quad \forall k .\label{keyeq}
\ee
Previous attempts of solving the above equations\cite{aniso,bio,caplan2} have
ended up with very complicated calculations and discussions. This is
because for a $N\times N$ Onsager matrix, there are $N(N+1)/2$ independent
response coefficients (if the Onsager matrix is symmetric). Besides,
there are $N-1$ coupled differential equations to solve
(from Eqs.~(\ref{c-f}) and (\ref{phi-def}), scaling all forces by a
constant does not change $\phi$; this property reduces the number of
differential equations to be solved by one). Solving these equations
analytically becomes a formidable task when $N\ge
3$ (see, e.g., the rather complicated discusssions in
Ref.~\cite{caplan2}). In this work we manage to solve the problem
analytically in a particularly simple way.

We notice that the force-current relation can be rewritten as
\begin{equation}
  \left( \begin{array}{c}
      \vec{{\cal J}}_O \\ \vec{{\cal J}}_{I} \end{array} \right) =
  \left( \begin{array}{cccc}
      \hat{{\cal M}}_{OO} & \hat{{\cal M}}_{OI} \\
      \hat{{\cal M}}_{IO} & \hat{{\cal M}}_{II}
    \end{array}\right) \left(\begin{array}{c}
      \vec{{\cal F}}_O \\ \vec{{\cal F}}_I \end{array}\right) , \label{oi-form}
\end{equation}
where the symbols $O$ and $I$ are used to abbreviate the indices of
forces and currents for exergy output and input, respectively. E.g.,
$\vec{{\cal J}}_O$ is the vector of the output current and $\hat{{\cal
    M}}_{OO}$ is the matrix relating the output force vector
$\vec{{\cal F}}_O$ to the output current vector $\vec{{\cal
    J}}_O$. Hence,
\begin{subequations}
  \begin{align}
    & \dot {\cal A}_{out}=-\vec{{\cal F}}_O^T \hat{{\cal M}}_{OI}\vec{{\cal
        F}}_I-\vec{{\cal F}}_O^{T}\hat{{\cal M}}_{OO} 
    \vec{{\cal F}}_O , \\ 
    & \dot {\cal A}_{in} = \vec{{\cal F}}_I^T
    \hat{{\cal M}}_{IO}\vec{{\cal F}}_O + \vec{{\cal F}}_I^{T}\hat{{\cal M}}_{II} 
    \vec{{\cal F}}_I , 
  \end{align} 
  \label{out-in} 
\end{subequations}
where the superscript $T$ stands for matrix (vector) transpose. For
symmetric Onsager matrix, $\hat{{\cal M}}_{II}=\hat{{\cal M}}_{II}^T$, 
$\hat{{\cal M}}_{OI}=\hat{{\cal M}}_{IO}^T$, and 
$\hat{{\cal M}}_{OO}=\hat{{\cal M}}_{OO}^T$. 

From Eqs.~(\ref{phi-def}), (\ref{keyeq}), and (\ref{oi-form}), we find that
\be
\partial_{\vec{{\cal F}}_O} \dot {\cal A}_{out} = \phi_{max}
( \partial_{\vec{{\cal F}}_O} \dot {\cal A}_{in} ) \label{cent}
\ee 
which gives
\be
\vec{{\cal F}}_O = - \frac{1+\phi_{max}}{2} \hat{{\cal M}}_{OO}^{-1} \hat{{\cal M}}_{OI}
\vec{{\cal F}}_I .
\ee
The inverse of the matrix $\hat{{\cal M}}_{OO}$ is justified as
$\hat{{\cal M}}_{OO}$ is a positive matrix. Inserting this into
Eq.~(\ref{phi}) we obtain
\begin{align}
& \phi_{max} =
\frac{\frac{1}{4}\left(1-\phi_{max}^2\right)\lambda}{1-\frac{1+\phi_{max}}{2}\lambda} \label{phimeq}
\end{align}
where $\lambda \equiv {\rm max}{\ave{\hat{\Lambda}}}$ and
$\ave{\hat{\Lambda}} \equiv \vec{{g}}^T {\hat{\Lambda}} \vec{{g}}$
with ${\vec g}$ being a normalized vector (i.e., ${\vec g}^T{\vec g}=1$) defined as
\be
{\vec g} \equiv \left . \hat{{\cal
      M}}_{II}^{1/2}\vec{{\cal F}}_{I}\right/\sqrt{  
\vec{{\cal F}}_I^T\hat{{\cal M}}_{II}\vec{{\cal F}}_{I}},
\ee
and
\begin{align}
& \hat{\Lambda}\equiv \hat{{\cal M}}_{II}^{-1/2}\hat{{\cal
    M}}_{IO}\hat{{\cal M}}_{OO}^{-1}
\hat{{\cal M}}_{OI}\hat{{\cal M}}_{II}^{-1/2} .
\label{matrix}
\end{align}
The inverse square root of the matrix $\hat{{\cal M}}_{II}$ is
well-defined since $\hat{{\cal M}}_{II}$ is a positive matrix
[see proof in Appendix~A]. 

Eq.~(\ref{phimeq}) is now a quadratic equation that can be solved
analytically. The physical solution with $\phi_{max}<1$ is
\be
\phi_{max} = \frac{\sqrt{\xi+1}-1}{\sqrt{\xi+1}+1}, \quad\quad \xi \equiv
\frac{\lambda}{1-\lambda} \label{phim}
\ee
where $\xi$ is the figure of merit and $\lambda$ is called the
``degree of coupling''\cite{caplan1}. We call the matrix
$\hat{\Lambda}$ as the ``coupling matrix''. Finally, $\vec{{\cal
    F}}_I$ or the normalized vector $\vec{{g}}$ must be tuned to
maximize $\ave{\hat{\Lambda}}$. The maximum value is achieved when
$\vec{{g}}$ equals to the eigenvector of $\hat{\Lambda}$ which
corresponds to the largest eigenvalue, which gives
\be
\lambda = {\rm largest\  eigenvalue\ of}\ \hat{\Lambda} . \label{phim2}
\ee
It is proven in Appendix~B that $\lambda\le 1$ as bounded by the
second law of thermodynamics. The $\lambda\to 1$ limit can be reached
only in the reversible limit when the determinant of the Onsager
matrix is zero\cite{note}. Eq.~(\ref{phim2}) represents one of the
main results in this work which was {\em not} found in
Ref.~\cite{caplan2} despite rather complicated treatment there.

The output power $\dot W=\dot {\cal A}_{out}$
at maximum exergy efficiency is
\be
\dot W (\phi_{max}) = \frac{1}{4}(1-\phi_{max}^2) \lambda \left(\vec{{\cal
    F}}_I^T\hat{{\cal M}}_{II}\vec{{\cal F}}_{I}\right) . \label{p-phim}
\ee

We now study the exergy efficiency for maximum power. The physical concern is
to optimize the output power by tuning the output forces 
$\vec{{\cal F}}_O$ which corresponds to adjusting the response of the
device accepting the output energy to maximize the output power
(as will be shown in the next section). The output power is then
optimized at $\partial_{\vec{{\cal F}}_O} \dot {\cal A}_{out}=0$ which renders
$\vec{{\cal F}}_O = - \frac{1}{2} \hat{{\cal M}}_{OO}^{-1} \hat{{\cal M}}_{OI}
\vec{{\cal F}}_I$.
The equation for $\phi$ can be established by inserting the above into
Eq.~(\ref{phi}) which is then solved in a way
similar to the solution of Eq.~(\ref{phimeq}). After that we optimize
$\phi$ by tuning the input forces $\vec{{\cal F}}_I$ and then obtain
the optimal exergy efficiency for maximum power as
\be
\phi_{opt}(\dot W_{max}) = \frac{ \xi }{2(\xi+2)} \le 50\% , \label{mpeff}
\ee
where $\xi$ is given in Eq.~(\ref{phim}) and $\lambda$ is again the largest
eigenvalue of the coupling matrix $\hat{\Lambda}$. The above
expression is consistent with the well-known result that
the upper limit of the exergy efficiency for maximum power for systems
with symmetric Onsager matrix is 50\%\cite{ca-ex,Seifert-review,odum,maxpower}. The
above derivations also provide a solid proof of the upper bound, 50\%,
for general thermodynamic systems in the linear-response regime.
The maximum output power is found to be
\be
\dot W_{max} =  \frac{1}{4} \lambda \left(\vec{{\cal
    F}}_I^T\hat{{\cal M}}_{II}\vec{{\cal F}}_{I}\right) . \label{mp}
\ee

\begin{figure}[htb]
  \includegraphics[height=4.5cm]{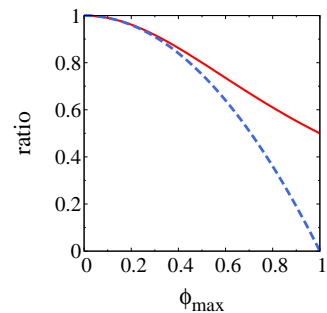}
  \caption{{(Color online) The ratio of optimal efficiency at maximum
    power to the maximum efficiency $\frac{\phi_{opt}(\dot
      W_{max})}{\phi_{max}}$ (solid curve) and the ratio of the power
    at maximum efficiency to the maximum power $\frac{\dot W
      (\phi_{max}) }{\dot W_{max} }$ (dashed curve) as functions of
    the maximum efficiency $\phi_{max}$, as given by
    Eq.~(\ref{eq:gen}), for thermodynamic machines with symmetric Onsager matrix.}}
\label{fig:rela}
\end{figure}

Comparing the exergy efficiencies and output powers for the two
optimization schemes discussed in this section, we find that
\begin{subequations}
\begin{align}
& \frac{\phi_{opt}(\dot W_{max})}{\phi_{max}} = \frac{1}{1+\phi_{max}^2},\\
& \frac{\dot W (\phi_{max}) }{\dot W_{max} } = 1-\phi_{max}^2 . 
\end{align}
\label{eq:gen}
\end{subequations}
{Remarkably, the above two simple relationships hold for all
thermodynamic machines with symmetric Onsager matrix in the
linear-response regime (thermodynamic systems with asymmetric
Onsager matrix is discussed in Sec.~\ref{sec:asym}). The above two
relationships bear very important information on the optimal
efficiencies and powers which is one of the main results in the
present work. Fig.~\ref{fig:rela} represents them graphically. Particularly 
in the reversible limit $\phi_{max}=1$, the output power at maximum
efficiency vanishes\cite{low-diss} while the efficiency at maximum power
reaches 50\% (These properties were proven to be general for
time-reversal symmetric systems in Ref.~\cite{Seifert-review} as
well). At low efficiency limit, $\phi_{max}\ll 100\%$, the power and
efficiency at the two optimal conditions are {\em almost the
  same}. Considerable differences between the two optimal conditions
appear only when $\phi_{max}\gtrsim 20\%$, $\xi\gtrsim 1$, or
$\lambda\gtrsim 0.5$.}

We remark that the largest eigenvalue of $\hat{{\cal
    M}}_{II}^{-1/2}\hat{{\cal M}}_{IO}\hat{{\cal M}}_{OO}^{-1}
\hat{{\cal M}}_{OI}\hat{{\cal
    M}}_{II}^{-1/2}$ is the same as the largest eigenvalue of 
$\hat{{\cal
    M}}_{OO}^{-1/2}\hat{{\cal M}}_{OI}\hat{{\cal M}}_{II}^{-1}
\hat{{\cal M}}_{IO}\hat{{\cal
    M}}_{OO}^{-1/2}$ (proof is given in
Appendix~B). Particularly in thermoelectric energy conversion, this
means that the figures of merit for the engine, refrigerator, and heat
pump are the same. 
These properties can be used to simplify the
calculation of the figure of merit when one of the two is easier to
calculate.

\begin{figure}[htb]
  \includegraphics[height=6.cm]{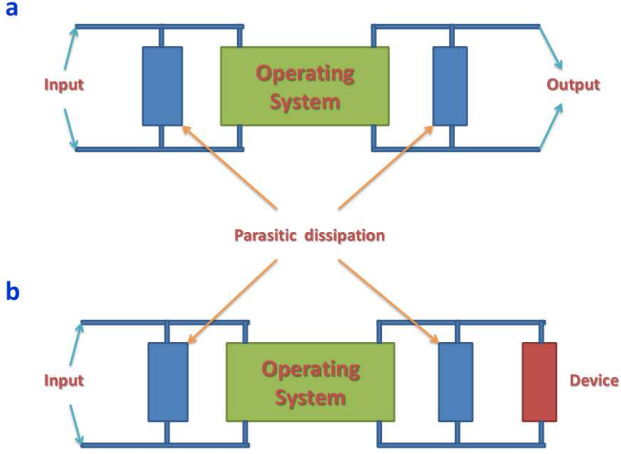}
  \caption{(Color online) Schematic of realistic thermodynamic
    machines. A machine accepts input energy and convert it into
    output energy. The output can be assigned to a huge reservoir
    (e.g., an electrical power grid with huge capacity) (a), or to a
    finite device (b). In realistic situations there are mechanisms
    that dissipate part of input energy and prevent it from 
    been converted into useful outputs, as well as mechanisms that
    consumes part of output energy and reduce the amount of useful
    outputs. These mechanisms are called ``parasitic dissipation''.}
\label{fig:illu}
\end{figure}

\section{Realistic considerations: output to a huge reservoir or to
  a finite device}
\label{real}

In realistic situations the input energy may pass through some
parallel channels without entering into the system which reduces the
amount of useful input energy. Besides, the output energy can also be
dissipated into channels parallel to the device accepting the output
power. These mechanisms are called as ``parasitic 
dissipation''. The effect is described by the following
phenomenological equations
\be
\vec{{\cal J}}_I^p = \hat{{\cal M}}_{II}^p \vec{{\cal F}}_I, \quad
\quad \vec{{\cal J}}_O^p = \hat{{\cal M}}_{OO}^p \vec{{\cal
    F}}_O .
\ee
Here the superscript $p$ stands for parasitic dissipation. The currents
for energy input into the operating system becomes $\vec{{\cal
    J}}_I+\vec{{\cal J}}_I^p$. And the currents that load into the
device becomes $\vec{{\cal J}}_O+\vec{{\cal J}}_O^p$.
The equivalent circuit is depicted in Fig.~\ref{fig:illu}. Taking into account of 
those parasitic currents modifies the response coefficients as
\be
\hat{{\cal M}}_{II}\to \hat{{\cal M}}_{II} + \hat{{\cal
    M}}_{II}^p,\quad 
\hat{{\cal M}}_{OO} \to \hat{{\cal M}}_{OO} +
\hat{{\cal M}}_{OO}^p .
\ee 
Parasitic dissipation increases the eigenvalues of the matrices
$\hat{{\cal M}}_{II}$ and $\hat{{\cal M}}_{OO}$ because both
$\hat{{\cal M}}_{II}^p$ and $\hat{{\cal M}}_{OO}^p$ are positive
matrices. As a consequence the degree of coupling $\lambda$ and the
figure of merit $\xi$ are reduced, according to Eqs.~(\ref{matrix})
and (\ref{phim}). This is consistent with the physical picture that
part of the useful energy is consumed by the parasitic dissipation.

Energy from the operating system can be outputted to (i) a huge
reservoir (e.g., a power grid with huge capacity), or to (ii) a finite
device. The optimization presented in Sec.~III is for option (i) where
the output current $\vec{{\cal J}}_O$ does not induce any observable
effect on the huge reservoir which in turn modifies the force 
$\vec{{\cal F}}_O$, so that $\vec{{\cal J}}_O$ and $\vec{{\cal F}}_O$
are uncorrelated. In electrical circuit analog, it is equivalent
to using the output energy to charge a huge capacitor where the charging
current $\vec{{\cal J}}_O$ does not change the voltage across the 
capacitor $\vec{{\cal F}}_O$. For option (ii) if the response of the
device is $\vec{{\cal J}}_O^{L} =  \hat{{\cal M}}_L \vec{{\cal F}}_O$,
the Kirchhoff's current law requires that $\vec{{\cal J}}_O +
\vec{{\cal J}}_O^{L} =0$. Therefore, 
\be
\vec{{\cal F}}_O 
= - (\hat{{\cal M}}_{OO}+\hat{{\cal M}}_{L})^{-1}\hat{{\cal
    M}}_{OI}\vec{{\cal F}}_I .
\ee 
The power consumed by the device is 
\bea
\vec{{\cal F}}_O^T\hat{{\cal M}}_L\vec{{\cal F}}_O&=&\vec{{\cal F}}_I^T\hat{{\cal M}}_{IO}(\hat{{\cal
  M}}_{OO}+\hat{{\cal M}}_L)^{-1} \hat{{\cal M}}_L \nn\\
&& \ \ \times (\hat{{\cal M}}_{OO}+\hat{{\cal
  M}}_L)^{-1}\hat{{\cal M}}_{OI}\vec{{\cal F}}_I.
\eea
The input exergy is
\be
\vec{{\cal F}}_I^T\vec{{\cal J}}_I=\vec{{\cal F}}_I^T(\hat{{\cal M}}_{II}-\hat{{\cal M}}_{IO}(\hat{{\cal
  M}}_{OO}+\hat{{\cal M}}_L)^{-1}\hat{{\cal M}}_{OI})\vec{{\cal F}}_I.
\ee 
The exergy efficiency is then
\be
\phi = \frac{\vec{{\cal F}}_O^T\hat{{\cal M}}_L\vec{{\cal
      F}}_O}{\vec{{\cal F}}_I^T\vec{{\cal J}}_I} .
\ee
{By varying $\hat{{\cal M}}_L$ of the device that receives power
  from the operating system,}
we find that the maximum output power is reached at
\be
\hat{{\cal M}}_L = \hat{{\cal M}}_{OO} , \label{mj}
\ee
whereas the maximum exergy efficiency is reached when
\be
\hat{{\cal M}}_L = \sqrt{1-\lambda} \hat{{\cal M}}_{OO} .
\ee
At these conditions we obtain again Eqs.~(\ref{phim}), (\ref{p-phim}),
(\ref{mpeff}), and (\ref{mp}). The above results reflect the
importance of matching between the response of the 
device $\hat{{\cal M}}_L$ and that of the system $\hat{{\cal M}}_{OO}$
in optimizing the efficiency and output
power\cite{imp-match}. {Particularly, Eq.~(\ref{mj}) generalizes the
maximum power theorem (Jacobi's Law for electrical circuits, i.e., ``Maximum power is
transferred when the internal resistance of the source equals the
resistance of the load, when the external resistance can be varied,
and the internal resistance is constant'') to all thermodynamic machines
with symmetric Onsager matrix in the linear-response regime.}

There are two possible schemes of adjusting the input forces, 
$\vec{{\cal F}}_I$, to optimize the performance of the machine. The
first scheme is to optimize the efficiency, i.e.,
to optimize $\lambda$. This has been discussed in Sec.~\ref{op1}. This
scheme reflects balance between optimizing output power and efficiency
which is relevant to some biological and ecological
systems\cite{odum}. The second scheme is to adjust $\vec{{\cal F}}_I$
for further optimization of the output power. This will lead to
efficiency smaller or equal to that in Eq.~(\ref{mpeff}). Hence the 
exergy efficiency for this scheme is also not larger than 50\%. From
Eqs.~(\ref{matrix}) and (\ref{mp}) one finds that
%\be
$\dot W_{max} = \frac{1}{4} \vec{{\cal F}}_I^T \hat{{\cal M}}_{IO} \hat{{\cal
  M}}_{OO}^{-1} \hat{{\cal M}}_{OI} \vec{{\cal F}}_I $.
%\ee
The above can be optimized to be
$\dot W_{max} = \frac{1}{4} \Upsilon \left(\vec{{\cal F}}_I^T
\vec{{\cal F}}_I\right)$,
with $\Upsilon$ being the largest eigenvalue of the matrix 
$\hat{{\cal M}}_{IO} \hat{{\cal M}}_{OO}^{-1} \hat{{\cal M}}_{OI}$.
It can be shown that $\Upsilon$ is positive [see Appendix~B]. There is
no obvious upper bound on it that is imposed by the laws of
thermodynamics (except maybe in the zero temperature limit\cite{joe}). The
above derivation is meaningful only when all input thermodynamic forces
${\cal F}_n$ ($\forall n\in I$) are {\em measured in the same physical
  unit and scale}. This requirement is usually not satisfied for
systems with more than one type of input forces (e.g., if both
mechanical and electrical forces are used for energy
input). Discussion on this scheme of performance optimization depends
on specific systems which is of little interest for our purpose.

\section{optimal exergy efficiency and power for systems with asymmetric
  Onsager matrix} 
\label{sec:asym}

We now study systems with asymmetric Onsager matrix. We first note
that $\vec{{\cal F}}_I^T\hat{{\cal M}}_{II}\vec{{\cal F}}_I={\cal
  F}_I^T\hat{{\cal M}}^s_{II}\vec{{\cal F}}_I$ and $\vec{{\cal F}}_O^T
\hat{{\cal M}}_{OO}{\cal F}_O=\vec{{\cal F}}_O^T\hat{{\cal M}}^s_{OO}
\vec{{\cal F}}_O$ where $\hat{{\cal
    M}}^s_{II}=\frac{1}{2}\left(\hat{{\cal M}}_{II}+\hat{{\cal
      M}}^T_{II}\right)$ and $\hat{{\cal
    M}}^s_{OO}=\frac{1}{2}\left(\hat{{\cal M}}_{OO}+\hat{{\cal
      M}}^T_{OO}\right)$. This property is due to the symmetry of the
summation over indices of forces.

It is hard to derive the optimal exergy efficiency and power for general
systems with asymmetric Onsager matrix {(see Appendix~C)}. Here we focus on a special
situation where
$\hat{{\cal M}}_{OI} = r \hat{{\cal M}}_{IO}^T$ with $r$ being a
real number. {Such a simplification
  is for the convenience of treatment instead of inspired by realistic
  physical systems.}
For this particular situation, from Eq.~(\ref{cent}), we
find
$\vec{{\cal F}}_O = - \frac{1+r^{-1}\phi_{max}}{2} \left(\hat{{\cal M}}_{OO}^{s}\right)^{-1} \hat{{\cal M}}_{OI}
\vec{{\cal F}}_I $.
Inserting this into Eq.~(\ref{phi}) and solving the equation for
$\phi_{max}$, we obtain
\be
\phi_{max} = r \frac{\sqrt{\xi+1}-1}{\sqrt{\xi+1}+1} \label{trbphiM}
\ee
where $\xi$ is given by the same expression as in Eqs.~(\ref{phim})
and (\ref{phim2}) but with $\hat{{\cal M}}_{OO}$ and $\hat{{\cal
    M}}_{II}$ replaced by 
their symmetric counterparts $\hat{{\cal M}}_{OO}^s$ and $\hat{{\cal
    M}}_{II}^s$. The exergy efficiency for maximum power is given by
\be
\phi_{opt}(\dot W_{max}) = \frac{r \xi}{2(\xi+2)} .
\ee
From the second law of thermodynamics the restriction on $\lambda$ is
[see Appendix~B]
\begin{subequations}
\begin{align}
& \frac{4r}{(1+r)^2}\le \lambda < 0 , \quad {\rm if}\quad
r < 0 , \\
& 0 \le \lambda \le \frac{4r}{(1+r)^2} , \quad {\rm if}\quad
r \ge 0 .
\end{align}
\label{ineq}
\end{subequations}
The above restrictions give rise to
$\xi + 1=\frac{1}{1-\lambda}\ge 0$ and $r(\sqrt{\xi +1} - 1) >0$,
so that the optimal exergy efficiency given in Eq.~(\ref{trbphiM}) is
positive and well-defined.

The maximum possible, i.e., the upper bound of exergy efficiency is
reached at $\lambda=\frac{4r}{(1+r)^2} \label{lam-zeta}$ as
\begin{subequations}
\begin{align}
& \phi_{bound} = r^2, \quad {\rm if}\quad |r| <
1 , \\
& \phi_{bound} = 1, \quad {\rm if}\quad |r| \ge 1 . \label{37b}
\end{align}
\label{phib}
\end{subequations}
The dissipation at the upper bound exergy efficiency is
\begin{subequations}
\begin{align}
& T\dot S_{tot} = (1-r)^2\left(\vec{{\cal F}}_I^T\hat{{\cal
      M}}_{II}^s\vec{{\cal F}}_I\right), \quad {\rm if}\quad |r| <
1 , \\
& T\dot S_{tot} = 0, \hspace{3.6cm} {\rm if}\quad |r| \ge 1 .
\end{align}
\end{subequations}
The entropy production for $|r| <1$ is always positive hence the
upper bound efficiency is not 100\%.

The upper bound of the exergy efficiency for maximum power is also reached at
$\lambda=\frac{4r}{(1+r)^2}$ with
\begin{align}
& \left.\phi_{opt}(\dot W_{max})\right|_{bound} = \frac{r^2}{r^2+1} .
\end{align}
From the above equation the Curzon-Ahlborn limit of exergy
efficiency\cite{ca,ca-ex,Seifert-review} $\phi_{CA}=50\%$ can be overcome when
$|r|>1$. This is first pointed out by Benenti {\sl et al.} in the
study of thermoelectric efficiency in systems with broken
time-reversal symmetry\cite{trb1}.

The output power at maximum exergy efficiency is 
\be
\dot W (\phi_{max}) = \frac{1}{4}(1-r^{-2}\phi_{max}^2) r\lambda \left(\vec{{\cal
    F}}_I^T\hat{{\cal M}}_{II}^s \vec{{\cal F}}_{I}\right) .
\ee
Combining the above with Eq.~(\ref{phib}), the upper bound of
efficiency for $|r|<1$ is $\phi=r^{2}$ so that the output
power is positive. For $|r|>1$ the maximum efficiency can reach
100\% without conflicting the requirement of positive output power.
The maximum output power is
\be
\dot W_{max} =  \frac{1}{4} r\lambda \left(\vec{{\cal
    F}}_I^T\hat{{\cal M}}_{II}^s\vec{{\cal F}}_{I}\right) .
\ee

We find that
\begin{subequations}
\begin{align}
& \frac{\phi_{max}}{\phi_{opt}(\dot W_{max})} = 1 + r^{-2}\phi_{max}^2 ,\\
& \frac{\dot W (\phi_{max}) }{\dot W_{max} }
= 1-r^{-2}\phi_{max}^2 . \label{41b}
\end{align}
\end{subequations}
Eqs.~(\ref{37b}) and (\ref{41b}) reveal that for systems with
asymmetric Onsager matrix with $|r|>1$, the output power is
nonzero even when $\phi_{max}$ reaches the value of 100\% in the
reversible limit. These results agree with the findings of Benenti
{\sl et al.} on thermoelectric efficiency and power in 
time-reversal symmetry broken systems\cite{trb1}. 

It is interesting to study the optimal exergy efficiency and power of the
reversed machine (i.e., the machine with output input reversed). The
output power of the reversed machine is $-\vec{{\cal F}}_I^T
\vec{{\cal J}}_I = - \dot {\cal A}_{in}$, while the input power becomes
$\vec{{\cal F}}_O^T \vec{{\cal J}}_O = - \dot {\cal A}_{out}$. The
reversed machine is working in the region with $\dot {\cal
  A}_I<0$. The efficiency of the reversed machine is defined as
\be
\phi^{\prime} = \frac{\dot {\cal A}_{in}}{\dot {\cal A}_{out}} .
\ee
We find that the optimal exergy efficiency and powers are similar but with
$r$ replaced by $r^{-1}$. Therefore for $|r|>1$ the
reversed machine can not reach the efficiency of 100\%, whereas for
$|r|<1$ the reversed machine can have 100\% efficiency with finite
power.

\begin{figure}[htb]
  \centerline{\includegraphics[height=4.2cm]{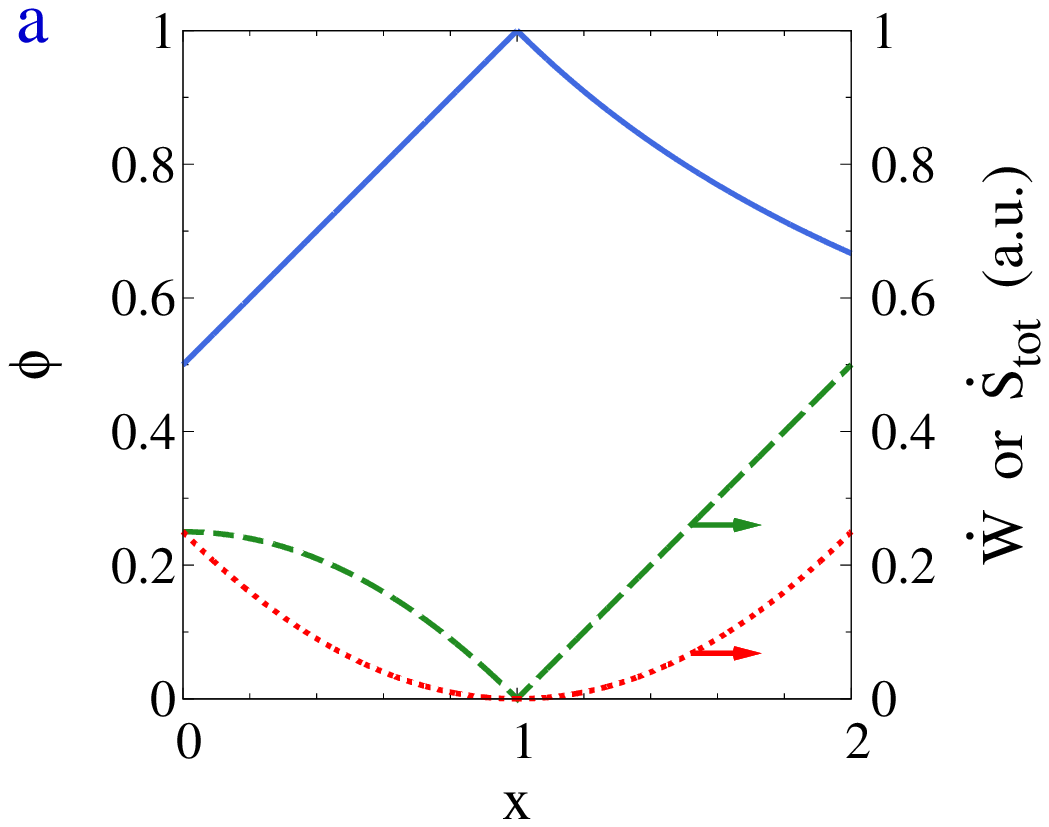}}
  \centerline{\includegraphics[height=4.2cm]{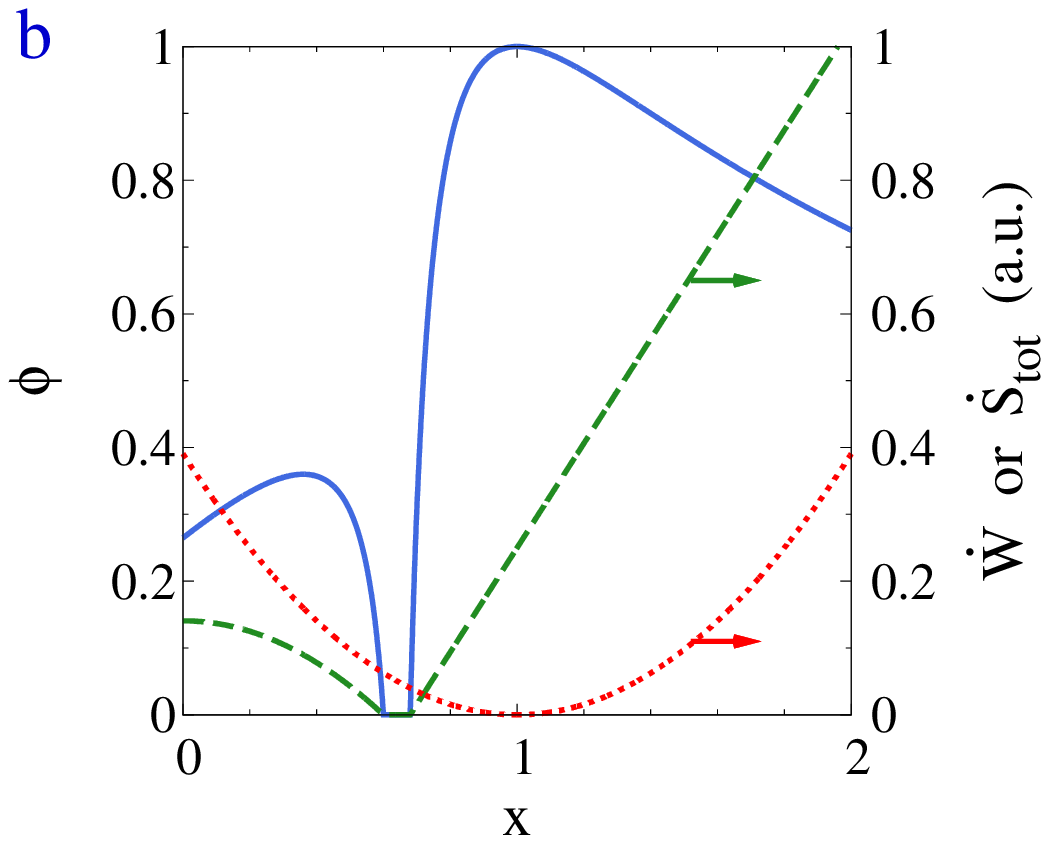}}
  \centerline{\includegraphics[height=4.2cm]{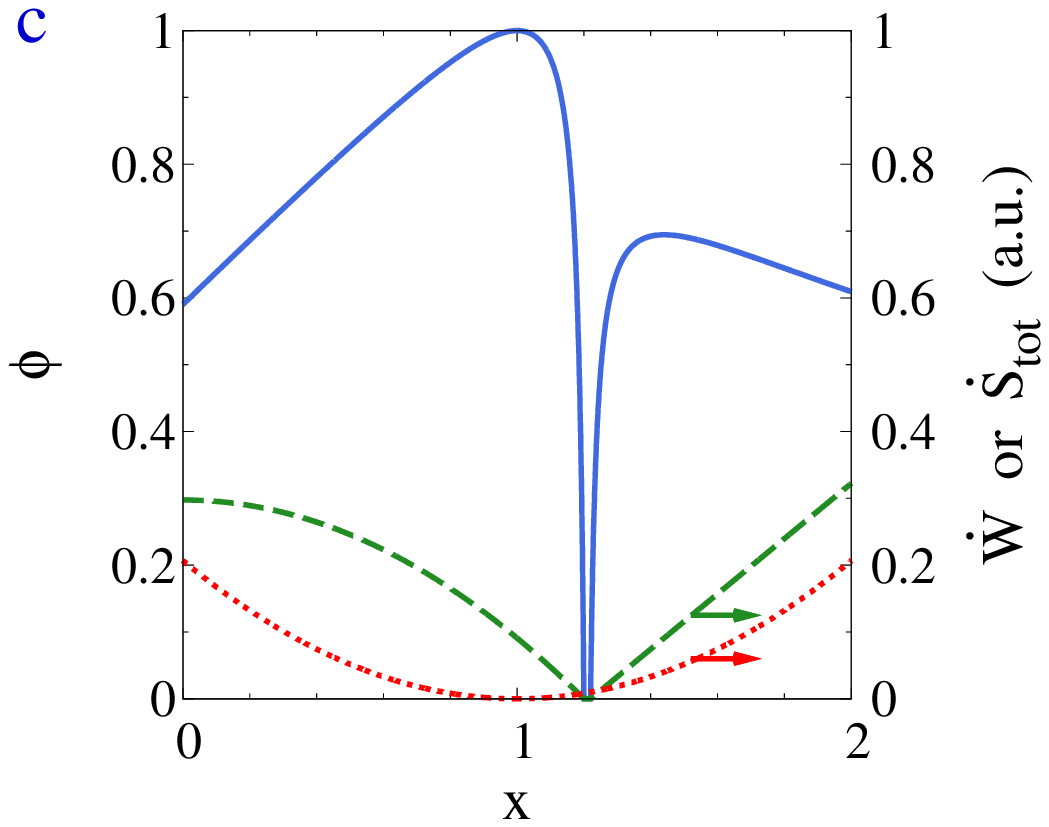}}
  \caption{ (Color online) The exergy efficiency $\phi$ (solid
    curves), output power $\dot W$ (dashed curves), and total entropy
    production $\dot S_{tot}$ (dotted curves) as functions
    of $x$ for $r=1$ (a), $r=0.6$ (b), and $r=1.2$
    (c). For each figure the left region with positive efficiency is
    the operating region of the machine, while the right region with
    positive efficiency is the operating region for the reversed
    machine. The definitions of efficiency and output power are
    different for the machine and the reversed machine.}
\label{fig:trb}
\end{figure}

To demonstrate this we plot the efficiency as a function of $x$ for 
$\vec{{\cal F}}_O=-\frac{r + x}{2r}\left(\hat{{\cal
      M}}_{OO}^s\right)^{-1}\hat{{\cal M}}_{OI}\vec{{\cal F}}_I$ in Fig.~\ref{fig:trb}. At
the limit with $\lambda=\frac{4r}{(1+r)^2}$ the efficiency is 
\be
\phi = \frac{r^2 - x^2}{1 + r^2 - 2 x } .
\ee
The output power $\dot W=\frac{r^2 - x^2}{(1+r)^2}(\vec{{\cal F}}_I^T\hat{{\cal
    M}}_{II}^s\vec{{\cal F}}_I)$ is positive when 
$|x|<|r|$. If $x>(1+r^2)/2$, both the input and output exergy
are negative which indicates that the machine is operating at the
reversed mode. The efficiency of the reversed machine is then 
\be
\phi^\prime = \frac{ 2 x - 1 - r^2  }{ x^2 - r^2}
\ee
The output power is $\dot
W=\frac{2x - 1 - r^2 }{(1+r)^2}(\vec{{\cal F}}_I^T\hat{{\cal
    M}}_{II}^s\vec{{\cal F}}_I)$. 

For all values of $r$ the reversible limit $T\dot
S_{tot}=0$ is reached at $x=1$. When $r=1$, 100\% efficiency is
reached by both the machine and the reversed machine at $x=1$ where
the input and output exergy as well as entropy production vanish [see
Fig.~\ref{fig:trb}a]\cite{low-diss}. For $|r|<1$, the machine cannot
reach to 100\% efficiency, but the reversed machine can reach 100\%
efficiency with finite output power, because at $x=1$ the machine is
operating in the reversed mode [see Fig.~\ref{fig:trb}b]. For $|r|>1$, the
output power of the machine is positive at $x=1$, thus the machine can
reach 100\% efficiency with finite output power [see Fig.~\ref{fig:trb}c].

In systems with broken time-reversal symmetry, such as two-dimensional
electron systems under perpendicular magnetic field, Hall effect, and
Nernst-Ettingshausen effect give rise to asymmetric Onsager
matrix\cite{chien,trb3}. The asymmetric Onsager matrix can be
decomposed into the symmetric part and anti-symmetric part. Specifically,
\be
\hat{{\cal M}}_{IO} = \hat{{\cal M}}_{IO}^s + \hat{{\cal M}}_{IO}^a, \quad \hat{{\cal M}}_{OI} = \hat{{\cal M}}_{OI}^s + \hat{{\cal M}}_{OI}^a
\ee
with $\hat{{\cal M}}_{IO}^s=(\hat{{\cal M}}_{OI}^s)^T$ and $\hat{{\cal
    M}}_{IO}^a=-(\hat{{\cal M}}_{OI}^a)^T$. The symmetric part,
$\hat{{\cal M}}_{IO}^s$, is related to entropy production and is
restricted by the second law of thermodynamics. The anti-symmetric
part, $\hat{{\cal M}}_{IO}^a$, however, does not contribute to
dissipation and is often related to Berry phase effects\cite{niu}.
The output and input exergy can be written as
\be
\dot {\cal A}_{out} = \dot {\cal A}_{out}^s - \vec{{\cal F}}_O^T
\hat{{\cal M}}_{OI}^a \vec{{\cal F}}_I, \quad \dot {\cal A}_{in} =
\dot {\cal A}_{in}^s - \vec{{\cal F}}_O^T \hat{{\cal M}}_{OI}^a
\vec{{\cal F}}_I \label{tr-de}
\ee
where $\dot {\cal A}_{out}^s$ and $\dot {\cal A}_{in}^s$ are the
output and input exergy for the symmetrized Onsager matrix with
\begin{align}
& \dot {\cal A}_{out}^s=-\vec{{\cal F}}_O^T \hat{{\cal M}}_{OI}^s\vec{{\cal
    F}}_I-\vec{{\cal F}}_O^{T}\hat{{\cal M}}_{OO}^s \vec{{\cal F}}_O , \nn\\
& \dot {\cal A}_{in}^s = \vec{{\cal F}}_I^T
\hat{{\cal M}}_{IO}^s\vec{{\cal F}}_O + \vec{{\cal F}}_I^{T}\hat{{\cal M}}_{II}^s 
\vec{{\cal F}}_I \nn .
\end{align} 
The additional term in Eq.~(\ref{tr-de}), $\vec{{\cal F}}_O^T
\hat{{\cal M}}_{OI}^a \vec{{\cal F}}_I$, does not cause entropy
production, but shift the input and output powers by the same
magnitude. In this way the reversible limit is shifted from the
boundary between the machine and the reversed machine, into the
operating region of the machine or the reversed machine, whichever has
positive output power in such limit.

{It should be emphasized here that although potential advantages of
  systems with asymmetric Onsager matrix have been predicted by
  Benenti {\sl et al.}\cite{trb1} from phenomenological theory (and
  extended in this work), no realistic physical system has been shown to
  have finite power at 100\% efficiency\cite{trb2,trb3}. It is very important to study
  efficiency and power of realistic physical systems with asymmetric Onsager
  matrix to clarify whether breaking time reversal symmetry could
  indeed improve the performance of a thermodynamic machine\cite{trb2,trb3}.}

\section{Application to realistic systems}
\label{exam}

\subsection{Example I: Thermoelectric energy conversion in isotropic
  systems}
Thermoelectric transport equation for an isotropic system is given by
\begin{equation}
\left( \begin{array}{c}
    \vec{j} \\ \vec{j}_{q} \end{array} \right) =
\left( \begin{array}{cccc}
    \sigma{\hat{\mathbf{1}}} & \sigma S T{ \hat{\mathbf{1}}}\\
    \sigma S T {\hat{\mathbf{1}}} & (\kappa T + \sigma S^2 T^2){\hat{\mathbf{1}}}
  \end{array}\right) \left(\begin{array}{c}
    \vec{{\cal E}} \\ - \vec{\grad} T / T \end{array}\right) ,
\end{equation}
where the electric field $\vec{{\cal E}}$ include both the external and induced electric
fields. Here $\sigma$ is the electrical conductivity, $S$ is the
Seebeck coefficient, $\kappa$ the thermal conductivity, {and
$\hat{\mathbf{1}}$ is the $3\times 3$ identity matrix}. The
efficiency, or coefficient of performance, of a thermoelectric refrigerator is 
\be
\eta \equiv \frac{\dot Q}{\dot W} = \frac{T}{\Delta T}  \frac{\vec{j}_q\cdot \vec{\grad}
  T/T}{\vec{j}\cdot \vec{{\cal E}}} = \eta_C \phi, \quad \eta_C \equiv
\frac{T}{\Delta T} . 
\ee
For a slab of thickness $\ell_z$ with temperature gradient and
electric field along the direction $z$ which is perpendicular to the
slab plane, the temperature difference is $\Delta
T=-\ell_z\frac{dT}{dz}>0$ for $\frac{dT}{dz}<0$. The maximum
coefficient of performance $\eta_{max}$ is related to the maximum exergy efficiency by
\be
\eta_{max} = \eta_C \phi_{max} = \eta_C
\frac{\sqrt{\xi+1}-1}{\sqrt{\xi+1}+1} . \label{cop-ex}
\ee
{The figure of merit is related to the degree of coupling which,
according to Eq.~(\ref{phim2}), is the largest eigenvalue of the
following coupling matrix
\be
\hat{\Lambda} = \frac{(\sigma S T)^2}{\sigma (\kappa T + \sigma S^2
  T^2)}\hat{\mathbf{1}} .
\ee
Since $\hat{\Lambda}$ is proportional to an identity matrix, the largest
eigenvalue is just 
\be
\lambda = \frac{\sigma S^2 T}{\kappa + \sigma S^2 T} .
\ee
Therefore the figure of merit is
\be
\xi = \frac{\lambda}{1-\lambda} = \frac{\sigma S^2 T}{\kappa} ,
\ee}
which recovers the well-known thermoelectric figure of merit as
found by Ioffe.

\subsection{Example II: Spin-thermoelectric effect}
\label{sec:s-th-cooling}
In conducting magnetic materials charge, spin, and thermal transports
are coupled together. There couplings are called spin-thermoelectric
or spin-caloric effect\cite{bauer}. In isotropic materials
spin-thermoelectric effect is described by the following transport
equation\cite{bauer} 
\be
\left( \begin{array}{c}
    \vec{j} \\ \vec{j}_s \\ \vec{j}_{q} \end{array} \right) =
\left( \begin{array}{cccc}
    \sigma{\hat{\mathbf{1}}} & \sigma P{\hat{\mathbf{1}}} & \sigma S T{\hat{\mathbf{1}}}\\
    \sigma P{\hat{\mathbf{1}}} & \sigma{\hat{\mathbf{1}}} & P^\prime \sigma S T{\hat{\mathbf{1}}} \\
    \sigma S T{\hat{\mathbf{1}}} & P^\prime \sigma S T{\hat{\mathbf{1}}} & \kappa_0 T{\hat{\mathbf{1}}}  
 \end{array}\right) \left(\begin{array}{c}
   \vec{{\cal E}} \\ - \vec{\grad} m  \\ -\vec{\grad} T / T \end{array}\right) .\label{s-te}
\ee
where $\vec{j}={\vec j}^{(\up)}+{\vec j}^{(\down)}$, $\vec{j}_s={\vec
  j}^{(\up)}-{\vec j}^{(\down)}$ with ${\vec j}^{(\up)}$ and ${\vec
  j}^{(\down)}$ denoting the electrical currents of the spin-up and
spin-down electrons, respectively. {$\vec{{\cal
    E}}=-\vec{\grad}\mu/e$ with $\mu\equiv (\mu_\up+\mu_\down)/2$},
and $m\equiv (\mu_\up-\mu_\down)/(2e)$ where $\mu_\up$ and $\mu_\down$ are
the electrochemical potentials for spin-up and spin-down electrons,
respectively, $e$ is the carrier charge. $\sigma$ is the electrical
conductivity, $S$ is the Seebeck coefficient, $P$ and $P^\prime$ are
two dimensionless quantities describing spin polarization of carriers
in different transport channels, $\kappa_0$ is the heat conductivity
at $\vec{{\cal E}}=\vec{\grad} m=0$. Microscopically they are given by
\begin{subequations}
\begin{align}
& \sigma = \int dE \left(-\frac{\partial n_F}{\partial E}\right) \sum_s
\sigma^{(s)}(E) ,\\
& P = \ave{s_z}, \quad S = \frac{\ave{E}}{eT} , \\
& P^\prime = \frac{\ave{E s_z}}{\ave{E}} ,\quad \kappa_0T =
e^{-2}\sigma \ave{E^2} .
\end{align}
\end{subequations}
with $\sigma^{(s)}(E)$ ($s=\up,\down$) being spin- and 
energy-dependent conductivity. We have set the energy zero to be at
the (equilibrium) chemical potential, i.e., $\mu\equiv 0$. $s_z=1$ or
-1 for spin up and down,
respectively. $n_F=1/[\exp(\frac{E}{k_BT})+1]$ is the 
Fermi distribution of the carrier. The averages in the above equations
are defined as
\be
\ave{{\cal O}} \equiv \sigma^{-1} \int dE \left(-\frac{\partial n_F}{\partial E}\right)
\sum_s \sigma^{(s)}(E) {\cal O}.
\ee
The above equations can be viewed as Mott relations\cite{MC}
generalized to spin-dependent transport. It assumes elastic transport
(by which the energy dependent conductivity is well-defined) and fails
when inelastic transport processes become important as pointed out by
the author and collaborators\cite{ourworks}. 

\begin{figure}[htb]
  \includegraphics[height=3.5cm]{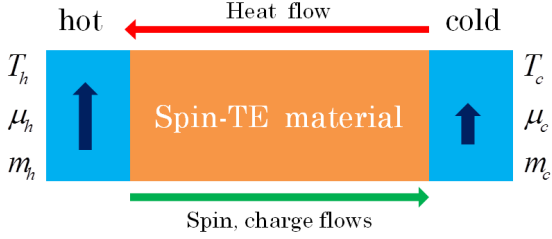}
  \caption{ (Color online) Spin-thermoelectric cooling. A
    spin-thermoelectric (``spin-TE'') material (i.e., a conducting 
    ferromagnetic material) sandwiched between two ferromagnetic
    electrodes with different temperature $T$, electrochemical
    potential $\mu\equiv (\mu_\up+\mu_\down)/2$, and
    spin accumulation $m\equiv (\mu_\up-\mu_\down)/(2e)$ where
    $\mu_\up$ and $\mu_\down$ are the electrochemical potentials for
    spin-up and spin-down electrons, respectively, and $e$ is the
    carrier charge. {For a set-up with $T_h>T_c$, $\mu_h>\mu_c$, and
      $m_h>m_c$ (the subscripts $h$ and $c$ denoting the hot and cold
      terminals, respectively), cooling (heat flowing from the cold terminal
      to the hot terminal) is driven by both the charge and spin flows.}}
\label{fig:tem1}
\end{figure}

We consider refrigeration driven by both the electric field
$\vec{{\cal E}}$ and the spin density gradient $\vec{\grad}m$.
The coefficient of performance of the refrigerator is defined as
\be
\eta \equiv \frac{\dot Q}{\dot W}= \frac{T}{\Delta T} \frac{\vec{j}_q\cdot \vec{\grad}
  T/T}{\vec{j}\cdot \vec{{\cal E}} - \vec{j}_s\cdot \vec{\grad}m} =
\eta_C \phi , \quad \eta_C = \frac{T}{\Delta T} .
\ee
Schematic of spin-thermoelectric cooling is shown in Fig.~\ref{fig:tem1}.
Consider a slab of thickness $\ell_z$ where the temperature
gradient, electric field, and spin density gradient are along the
direction perpendicular to the slab plane, i.e., the $z$ direction.
The temperature difference is $\Delta T = -\ell_z\frac{dT}{dz}>0$ for
$\frac{dT}{dz}<0$. The maximum coefficient of performance is again
related to the maximum exergy efficiency as given in
Eq.~(\ref{cop-ex}). Using Eqs.~(\ref{phim}) and (\ref{s-te}) we obtain
\be
\xi = \frac{\sigma T S^2(1-2PP^\prime+P^{\prime 2})}{\kappa_0
  (1-P^2) - \sigma T S^2(1-2PP^\prime+P^{\prime 2})} . \label{xi-tem}
\ee
Remarkably one can show that the above degree of coupling is {\em
  greater} than the figure of merit for thermoelectric cooling,
\be
\xi_{TE} = \frac{\sigma T S^2}{ \kappa_0 - \sigma T S^2 } ,\label{xi-te}
\ee
and the figure of merit for spin-Peltier cooling\cite{bauer,sp},
\be
\xi_{SP} = \frac{\sigma T S^2 P^{\prime 2}} { \kappa_0 - \sigma T S^2
  P^{\prime 2}} .\label{xi-sp}
\ee

This interesting phenomenon has a geometric origin which is understood
as follows. The electric field and the spin-density gradient can be 
parametrized as 
\be
\vec{{\cal E}} = \vec{{\cal F}}_0\cos\theta, \quad - \vec{\grad} m = \vec{{\cal
  F}}_0\sin\theta
\ee
where $\vec{{\cal F}}_0=\vec{{\bf e}}_z
\sqrt{\frac{1}{2e^2}(|\vec{\grad}\mu_\up|^2+|\vec{\grad}\mu_\down|^2)}$
with $\vec{{\bf e}}_z$ being the transport direction. $|\vec{{\cal F}}_0|$
is the total ``magnitude'' of the input force. The heat current,
\be
\vec{j}_q = \vec{j}_{q0} + \vec{j}_{q1} + \vec{j}_{q2} ,
\ee
consists of three parts: thermal conduction
$\vec{j}_{q0}=-\kappa_0\vec{\grad}T$, Peltier cooling $\vec{j}_{q1}=\sigma
ST\vec{{\cal E}}$, and spin-Peltier cooling
$\vec{j}_{q2}=-P^\prime\sigma ST\vec{\grad}m$. 
The cooling is achieved when the sum of the Peltier current
$\vec{j}_{q1}$ and the spin-Peltier current $\vec{j}_{q2}$ exceeds the
thermal conduction current $\vec{j}_{q0}$.

\begin{figure}[htb]
  \includegraphics[height=5.4cm]{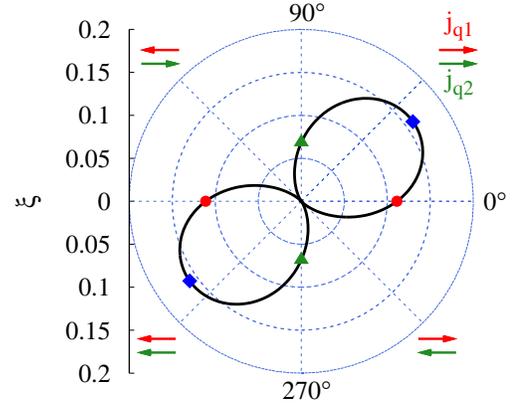}
  \caption{ (Color online) Polar plot of $\xi$ vs. $\theta$. The
    parameters are {$P=0.2$, $P^\prime=0.8$}, $S=50$~$\mu$V/K, and
    $T=300$~K. The heat 
    conductivity is $\kappa_0=\sigma L T$ with the Lorenz number of
    $L=2.5\times 10^{-8}$~W~${\rm \Ome}$~K$^{-2}$. The arrows indicates
    the {\em relative} direction between $\vec{j}_{q1}$ (red arrows) and
    $\vec{j}_{q2}$ (green arrows). The red dots represent the thermoelectric figure
    of merit $\xi_{TE}$, the green triangles represent the spin-Peltier
    figure of merit $\xi_{SP}$, and the blue rhombuses denote the
    figure of merit $\xi$ of combined thermoelectric and spin-Peltier cooling.}
\label{fig:tem}
\end{figure}

Tuning the angle $\theta$ changes the relative amplitude of the
Peltier and spin-Peltier heat currents, $\vec{j}_{q1}$ and
$\vec{j}_{q2}$. These two currents can be of the same sign, or the
opposite sign, depending on $\theta$. When  $\vec{j}_{q1}$ and
$\vec{j}_{q2}$ have the same sign, the cooling is enhanced, leading to
higher efficiency. However, when $\vec{j}_{q1}$ and
$\vec{j}_{q2}$ have opposite sign, the cooling is suppressed and the
efficiency is reduced. This is explicitly shown in Fig.~\ref{fig:tem}.
The underlying physics is more complicated when the input
work $\dot W$ is taken into consideration as well. However, this
simplified picture gives a snapshot that the two cooling mechanisms
can have cooperative effects.

\begin{figure}[htb]
  \includegraphics[height=3.8cm]{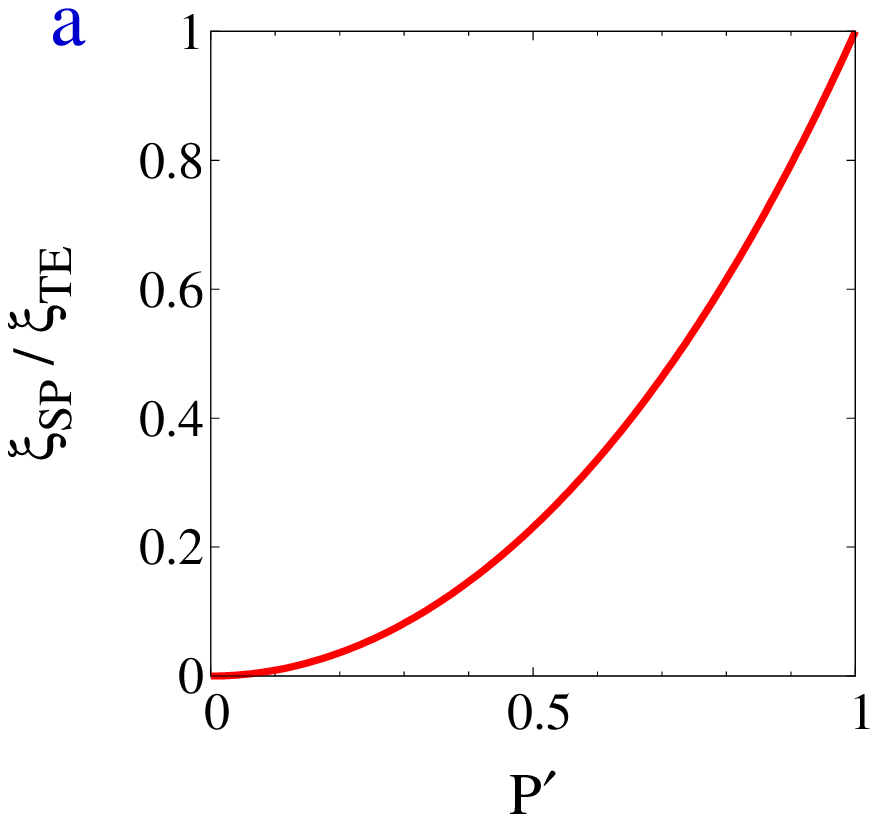}\includegraphics[height=4.1cm]{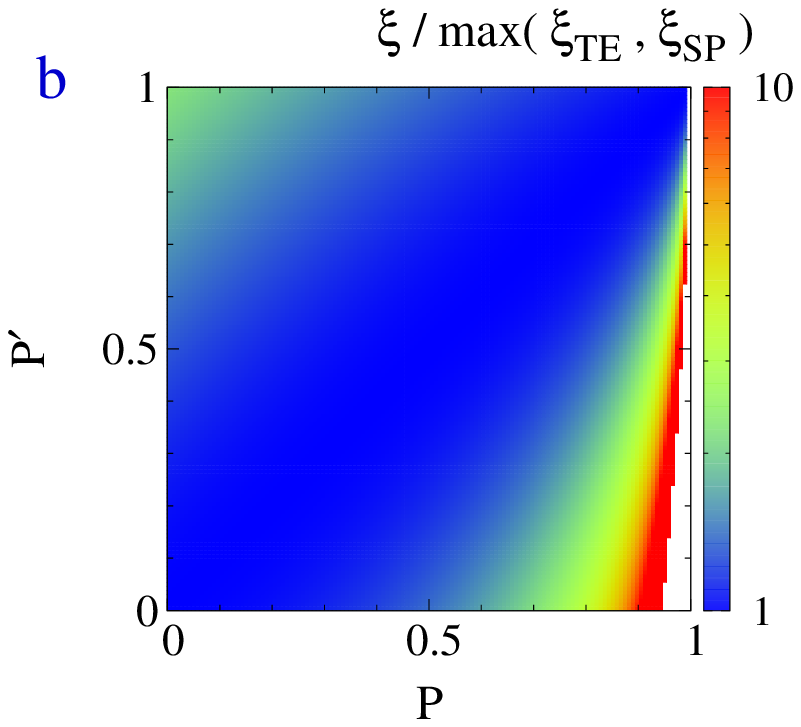}
  \caption{ (Color online) (a) The ratio of the figure of merit of
    spin-Peltier cooling $\xi_{SP}$ to that of thermoelectric cooling
    $\xi_{TE}$ as a function of $P^\prime$. (b) The enhancement of
    figure of merit due to cooperative effect, $\xi/{\rm
      max}(\xi_{TE},\xi_{SP})$, as a function of $P$ and $P^\prime$. The
    parameters are $S=50$~$\mu$V/K and $T=300$~K. The heat
    conductivity is $\kappa_0=\sigma L T$ with the Lorenz number of
    $L=2.5\times 10^{-8}$~W~${\rm \Ome}$~K$^{-2}$. The white region in
    (b) near $P=1$ is forbidden by the second law of thermodynamics.}
\label{fig:tem3}
\end{figure}

We also calculated the figure of merit for spin-Peltier cooling
$\xi_{SP}$ as function of $P^\prime$ according to Eq.~(\ref{xi-sp}) as
shown in Fig.~\ref{fig:tem3}a for $\sigma S^2 T/\kappa_0=0.1$. For the
same parameter, we plot the enhancement factor $\xi/{\rm
  max}(\xi_{TE},\xi_{SP})$ as function of $P$ and $P^\prime$ in
Fig.~\ref{fig:tem3}b. Significant enhancement of figure of merit due
to cooperative effect is attainable when $P^\prime$ deviates from $P$
markedly.

Efficient spin-thermoelectric cooling demands large Seebeck
coefficient. According to the literature, large Seebeck coefficient
ranging from 100 to 45000~$\mu$V/K can be attained in magnetic or
strongly-correlated semiconductors\cite{mag-sem} and magnetic tunnel
junctions\cite{mtj}. Sizable figure of merit, $\xi\sim 1$, however, is
still to be achieved\cite{mag-sem}.

The figure of merit at fixed $\theta$ is found as
\be
\xi(\theta) = \frac{\sigma S^2 T (P^\prime \sin\theta +
  \cos\theta)^2(1+2P\sin\theta\cos\theta)}{\kappa_0 - \sigma S^2 T (P^\prime \sin\theta +
  \cos\theta)^2(1+2P\sin\theta\cos\theta) }
\ee
The maximum exergy efficiency is achieved at $\theta=\theta_M$ with
\be
\tan\theta_M = \frac{P^\prime - P}{ 1 - P^\prime P} .
\ee
The figure of merit at $\theta=\theta_M$ is exactly the same as that
given in Eq.~(\ref{xi-tem}), which is greater than the
figures of merit for thermoelectric and spin-Peltier cooling,
$\xi_{TE}$ and $\xi_{SP}$, unless $P=P^\prime$. Such cooperative
effect prevails in systems with multiple cross-correlated responses,
which can be exploited to improve the efficiency. The discussions here
can also be applied to the efficiency and figure of merit of
spin-thermoelectric power generators\cite{tem-gen}.

\subsection{Example III: Piezoelectric, piezomagnetic and
  magnetoelectric effects}
\label{sec:p-e-m}

Piezoelectric energy harvest has been studied extensively and made
into useful devices\cite{pe-review}. There is also the piezomagnetic
effect where elastic strain induces a magnetization or vice
versa\cite{tem1}. These two effects are common in ferroelectric and 
ferromagnetic insulators\cite{tem1}. Materials with simultaneous
ferroelectric and ferromagnetic properties, or more generally multiple
spontaneous electric and magnetic orders\cite{tem1,multi1,science},
are called multiferroics. An important technologically property of
multiferroics is the magnetoelectric effect which
offers efficient conversion between electric and magnetic energy in
the radio frequency regime\cite{tem1}. Wood and Austin\cite{wood}
suggested many possible applications of the magnetoelectric effect,
among which there are transducers which convert the microwave magnetic
field into microwave electric field, attenuators which are used to
improve impedance matching in circuits, and ultrasensitive magnetic
field sensors\cite{tem1}. Multiferroics with strong magnetoelectric
response have been the aim of extensive studies\cite{tem1}. Recently,
strong magnetoelectric response were found in both
crystalline (such as CaMn$_7$O$_{12}$\cite{crys1},
TbMnO$_3$\cite{crys2}, and HoMnO$_3$\cite{crys3}) and nano-composite
(such as BiFeO$_3$ thin film heterostructures\cite{comp1} and
BaTiO$_3$-CoFe$_2$O$_4$ nano-structures\cite{comp2}) materials. In many of these
materials the interplay of piezoelectric and piezomagnetic responses
play an important role. In fact, multiferroics can be made from
nano-composites of ferroelectric and ferromagnetic compounds where
elastic strain at interfaces mediate coupling between electric
and magnetic polarizations\cite{science,nan}.

In these materials a full description of responses to external
mechanical, electric, and magnetic forces are given by\cite{nan,tem1}
\be
\left( \begin{array}{c}
    \hat{S} \\ \vec{D} \\ \vec{B} \end{array} \right) =
\left( \begin{array}{cccc}
    \hat{s} & \hat{d} & \hat{q} \\
    \hat{d}^T & \hat{\ep} & \hat{\alpha} \\
    \hat{q}^T & \hat{\alpha}^T & \hat{\mu}_m
 \end{array}\right) \left(\begin{array}{c}
   \hat{T} \\ \vec{E}  \\ \vec{H} \end{array}\right) .
\ee
where the forces are stress $\hat{T}$, electric field $\vec{E}$, and
magnetic field $\vec{H}$, the currents are strain $\hat{S}$, electric
displacement $\vec{D}$, and magnetic induction $\vec{B}$. Here
$\vec{D}$ and $\vec{B}$ stand for the values deviate from the
equilibrium ones (which could be nonzero in materials with spontaneous
polarization and magnetization). The response matrix has the dimension
of $12\times 12$. Specifically, $\hat{s}$ is the $6\times 6$
compliance tensor, $\hat{\ep}$ is the $3\times 3$ dielectric tensor,
$\hat{\mu}_m$ is the ($3\times 3$) permeability tensor, $\hat{d}$
describes piezoelectric response, $\hat{q}$ describes piezomagnetic
response, and $\hat{\alpha}$ gives magnetoelectric response. 

In general the response matrix is frequency dependent. Experiments
have shown resonance behavior in magnetoelectric response\cite{reson}.
Without further complication of specific circuits set-up for energy
conversion at finite frequencies\cite{pe1,pe2}, here we consider the
low-frequency limit which is sufficient to demonstrate the underlying
principles. Extension of study to finite frequency regimes
will be achieved in future works. First, the coupling matrix for
piezoelectric energy conversion is
\be
\hat{\Lambda}_{pe} = \hat{\ep}^{-1/2}~ \hat{d}^T~
\hat{s}^{-1} ~ \hat{d}~ \hat{\ep}^{-1/2} ,
\ee
which coincides with the ``electromechanical coupling tensor'' introduced in
Ref.~\cite{pe}. The largest electromechanical coupling factor of a
material is given by the largest eigenvalue of the coupling matrix
$\hat{\Lambda}_{pe}$. Piezoelectric effect allows harvest of mechanical
energy to power portable and isolated electrical systems as well as
small motors which have already found applications\cite{pe-review}.
Existing materials have already shown large electromechanical coupling
factors, reaching to $\gtrsim 0.5$\cite{pe1,ryu}, which allows efficient
piezoelectric energy conversion. In realistic systems, additional
mechanical and electrical damping reduces the efficiency\cite{pe1,pe2}.
Although further complication must be considered for a finite
frequency set-up with a mechanical oscillator, the efficiency is still
an increasing function of the electromechanical coupling
factor\cite{pe1,pe2}. Piezomagnetic effect can be used
for magnetic field sensing, stress sensing, and mechanical generation
of spin-waves\cite{tem1}. The coupling matrix for piezomagnetic energy
conversion is
\be
\hat{\Lambda}_{pm} = \hat{\mu}_m^{-1/2}~ \hat{q}^T~
\hat{s}^{-1} ~ \hat{q}~ \hat{\mu}_m^{-1/2} .
\ee
The largest piezomagnetic coupling factor is the largest eigenvalue of
the above matrix. Piezomagnetic coupling factor can be as large as 0.5
as well\cite{dong}. The coupling matrix for magnetoelectric energy
conversion is
\be
\hat{\Lambda}_{em} = \hat{\mu}_m^{-1/2}~ \hat{\alpha}^T~
\hat{\ep}^{-1} ~ \hat{\alpha}~ \hat{\mu}_m^{-1/2} .
\ee
Experiments on laminated composites of rare-earth-iron alloys
(Terfenol-D) and lead-zirconate-titanate (PZT) achieved a
magnetoelectric coefficient along the stacking direction as high as 
$\alpha_E=\alpha/\ep=$10~V~cm$^{-1}$~Oe$^{-1}$\cite{ryu}. Along this
direction the relative dielectric constant is about 1000\cite{ryu} and
the relative permeability is about 4\cite{nan2}. According to these
parameters, the magnetoelectric coupling factor along the stacking
direction is around 0.1. The largest magnetoelectric coupling factor is
given by the largest eigenvalue of the matrix $\hat{\Lambda}_{em}$.

The system also allows multiple input or output energy forms. For
example, magnetic energy can be generated by simultaneously inputting
electric and mechanic energy. This yield the coupling matrix of
\be
\hat{\Lambda}_{m-pe} = \hat{\mu}_m^{-1/2}~ \hat{q}_{pe}^T~
\hat{h}_{pe}^{-1} ~ \hat{q}_{pe}~ \hat{\mu}_m^{-1/2} 
\ee
where
\be
\hat{q}_{pe} = \left( \begin{array}{c}
    \hat{q} \\ \hat{\alpha} \end{array} \right), \quad  \hat{h}_{pe} = \left( \begin{array}{cccc}
    \hat{s} & \hat{d}  \\
    \hat{d}^T & \hat{\ep} 
 \end{array}\right) .
\ee
Similar to the results in Sec.~\ref{sec:s-th-cooling}, cooperative
effect will lead to larger degree of coupling from the above coupling
matrix. That is, the exergy efficiency is {\em no less} than those of
piezomagnetic effect and magnetoelectric effect. Significant
improvement of efficiency could be possible by the synergetic effect
in systems with cross-correlated piezo-electric-magnetic effect.

\subsection{Example IV: Biological energy conversion}

\begin{figure}[htb]
  \includegraphics[height=4cm]{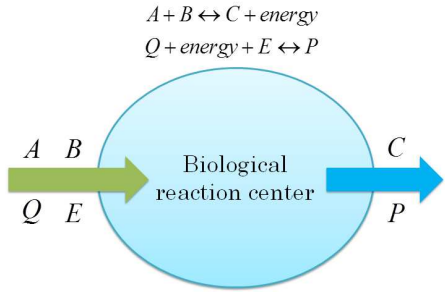}
  \caption{(Color online) Energy conversion in biological
    reaction. Biological reactions, $A+B\leftrightarrow C + energy$
    and $Q+energy+E\leftrightarrow P$, take place in the reaction
    center. The first reaction produces energy which is stored in
    material $P$ via the second reaction. At steady states there are
    continuous flows of materials across the membrane of the reaction
    center to facilitate continuous reactions. The membrane keeps a
    density (chemical potential) difference between the reaction
    center and the outside to control reaction rates. Arrows in the
    figure indicate possible flows of materials when energy is
    produced and stored in $P$.}
\label{fig:bio}
\end{figure}

Biological processes are driven by various energies: the internal
energy produced by oxidation and external energy from environments.
Understanding of bioenergetics is one of the most important and
challenging task in biology. Many of the processes can be described by
Onsager's linear-response theory (although many others
cannot)\cite{caplan1,caplan2,bio1,bio2,bio3}. One example is transport
across a membrane. The flows of various ions, such as Na$^+$,
Ca$^{2+}$, and H$^+$ as well as other materials, such as
phosphorylation, oxygen, and sugars are all driven by their density
gradients, chemical reaction and other forces\cite{bio}. If, e.g.,
some of these materials involve in a chemical reaction, flows of those
materials will be correlated. Synergetic effects will appear as 
multiple flows take place in coorperative ways. Biological systems,
may also utilize the cross-correlation of those flows to optimize
energy efficiency. There have been a lot of studies of bioenergetics
using irreversible
thermodynamics\cite{bio1,bio2,bio3,caplan1,caplan2}. However, none of 
them have reached a simple analytic results as obtained in this work.

To demonstrate the usefulness of the theory, we consider a toy model
describes the reaction of
\be
A + B \leftrightarrow C + energy, \quad  Q + energy + E \leftrightarrow P
\ee
in a reaction center surrounded by a membrane. We assume the reactions
are reversible with the help of enzymes. In the former reaction
$A$ and $B$ are consumed to produce $C$ while some energy is generated
which is absorbed by $Q$ and $E$ to form $P$ (energy stored in
$P$). We assume that all energy generated in the former reaction is
absorbed by the latter one. To describe such a reaction, we use
six flows, $J_A$, $J_B$, $J_Q$, and $J_E$ to describe the rate of
consumption of $A$, $B$, $Q$, and $E$, $-J_C$ and $-J_{P}$ to
describe the rate of production of $C$ and $P$. The flow and reaction
is illustrated in Fig.~\ref{fig:bio}. The reaction is
described by Eq.~(\ref{oi-form}) in linear-response regime with
\begin{subequations}
\begin{align}
& \vec{\cal J}_I = ( {\cal J}_A, {\cal J}_B, {\cal J}_Q, {\cal J}_E)^T , \quad \vec{\cal J}_O = (
{\cal J}_C, {\cal J}_P )^T , \\
& \vec{\cal F}_I = ( {\cal F}_A, {\cal F}_B, {\cal F}_Q, {\cal F}_E)^T
, \quad \vec{\cal F}_O = ( {\cal F}_C, {\cal F}_P )^T .
\end{align}
\end{subequations}
The forces can be written as ${\cal F}_i=\delta\mu_i + a_i$ where
$\delta \mu_i=\mu_i^{out}-\mu_i^{in}$ with $\mu_i^{out}$ and
$\mu_i^{in}$ being the chemical potential of $i$ outside and inside
the reaction center, respectively, $a_i$ is the affinity of material
$i$ for the reaction which is the free energy of $i$ per mole
(if ${\cal J}_i$ is measured in unit of mole per second). Biological
systems can control those flows and their correlations through
chemical reaction processes (e.g., via enzymes) as well as selective
and tunable transmission of materials through the membrane.
The efficiency of the biological reaction is 
$\phi=-\vec{\cal
  F}_O^T\vec{\cal J}_O/(\vec{\cal F}_I^T\vec{\cal J}_I)$.
The optimal efficiency is then given by
Eq.~(\ref{phim}) where the degree of coupling is given by the largest
eigenvalue of the coupling matrix $\hat{\Lambda}$ given by
Eq.~(\ref{matrix}). This result is much simpler than that discussed in
Ref.~\cite{caplan2}.

\section{Conclusion and discussions}

{We examined the important question of ``what is the maximum efficiency of a
thermodynamic machine when its linear responses to the external is
given?''. This question has been answered in simple limits with two
thermodynamic currents. It becomes rather difficult to answer for
a thermodynamic machine with arbitrarily complex responses. Efforts on
the problem in the literature failed in yielding general and analytic
results that are useful for material and structure engineering in
advanced energy technologies. Pushed by fast developing nanotechnology
and material technologies, complex systems with advanced functions
play more and more important roles. It becomes increasingly demanded to
extend the known, simple results on efficiency optimization with two
thermodynamic currents to those complex systems which is characterized
by a $N\times N$ Onsager matrix ($N>2$).}

We derived the optimal efficiency and powers
for general thermodynamic machines with arbitrary linear-response
coefficients. The results are written in simple and analytic forms.
Based on those results we establish two general relationships between
the optimal efficiency and powers for two realistic optimization
schemes: (i) maximum efficiency and (ii) optimal efficiency for
maximum power. We proved that the upper bound
efficiency at maximum output power is 50\% for all thermodynamic
systems with symmetric Onsager response matrix. The results are
confirmed by considering realistic energy systems where the output
power is consumed by a device of which the response coefficients
can be varied. {We proved that the maximum output power is reached when
the response matrix of the device receiving the power, $\hat{{\cal
    M}}_L$, is equal to that of the power-supplying machine in the
output sector, $\hat{{\cal M}}_{OO}$. This proof generalizes the
maximum power theorem (Jacobi's Law) to all thermodynamic machines
with symmetric Onsager matrix in the linear-response regime.} We also
extend the studies to systems with asymmetric Onsager matrix {(for
  a particular class of systems)} where the
efficiency at maximum output power can exceed 50\%. {Besides, in such
systems the second law of thermodynamics does not forbid the
reversible limit of efficiency, 100\%, to be reached at {\em finite}
output power. This phenomenon is caused by redistribution of free
energy between the input and output channels induced by
dissipationless responses (e.g., by magnetic field, geometric phases,
etc). We also show that such limit can only be reached in a machine by
its normal mode or reversed mode, but not by both of them.}

Several examples are presented to demonstrate applications of the
theory. First for isotropic thermoelectric systems, we recover Ioffe's
well-known results. We then consider refrigeration in
spin-thermoelectric systems. It is shown that driving cooling by both
electrochemical potential and spin density gradients yield maximum
efficiency {\em considerably higher} than when only one of the two gradients
(forces) is applied to the system. Such enhancement of maximum
efficiency due to cooperative effects between different forces
can be significant in certain parameter regimes. We remark that such
cooperative effects prevail in systems with multiple cross-correlated
responses and can be used to improve energy efficiency for realistic
machines. We also apply the theory to discussions of piezoelectric,
piezomagnetic, and magnetoelectric energy conversion and their
cooperative effects as well as biological energy conversion.
Studies in this work shed light on general properties of optimization
in energy applications and are helpful in guiding the search for high
performance energy materials and systems.

\section*{Acknowledgements}
I am greatly indebted to Rashmi C. Desai for a lot of discussions and
encouragements. I also wish to thank Yoseph Imry, Ora Entin-Wohlman,
Sajeev John, Christian van den Broeck, Baowen Li, Ming-Qi Weng, Gang Chen, Sidhartha Goyal,
Chushun Tian, and Daoyong Chen for illuminating discussions and
comments. This work was supported by the NSERC of Canada, the Canadian Institute for Advanced
Research, and the United States Department of Energy Contract
No. DE-FG02-10ER46754. Special thanks to CPTES at
Tongji University and IAS at Tsinghua University for hospitality where
parts of this work were completed.

\begin{appendix}

\section{Positiveness of Onsager matrix and definition of inverse
  square root of matrices}

The second law of thermodynamics requires $\dot S_{tot}\ge 0$ for 
all possible values of forces. That is
\bea
T\dot S &=& \sum_{nk} {\cal F}_n {\cal M}_{nk} {\cal F}_k \ge 0, \quad
\forall \vec{{\cal F}} ,\nn\\
&=& \sum_{nk} {\cal F}_n {\cal M}_{nk}^s {\cal F}_k \ge 0, \quad
\forall \vec{{\cal F}} ,
\eea
where ${\cal M}^s_{nk}=\frac{1}{2}\left({\cal M}_{nk}+{\cal
    M}_{kn}\right)$. Since $\hat{{\cal M}}^s$ is a real symmetric
matrix with dimension $N\times N$, it has $N$ (real) eigenvectors and
eigenvalues. For any vector $\vec{{\cal F}}$ can be decomposed into
the eigenvectors,
\be
\vec{{\cal F}} = \sum_{i=1}^{N} f_i \vec{{e}}_i ,
\ee
with $\vec{{e}}_i$ corresponding to the eigenvalue $m_i$, then
\be
T\dot S = \sum_i m_i f_i^2 .
\ee
The above is positive definite only when $m_i\ge 0$ for all $i$.
That is, all eigenvalues of the matrix $\hat{{\cal M}}^s$ must be
positive (In this work we take the situation with $m_i=0$ as the limit
that is approached from the $m_i>0$ side, which has never been reached
in realistic systems).

When $\hat{{\cal M}}_{II}$ is a real symmetric matrix there always exist
an orthogonal matrix $\hat{\Ome}_I$ such that $\hat{{\cal M}}_{II}
=\hat{\Ome}_I^T\hat{D} \hat{\Ome}_I$ where $\hat{D}$
is a diagonal matrix. According to the second law of thermodynamics
all the eigenvalues of matrix $\hat{{\cal M}}_{II}$ are positive.
Therefore all the elements of the diagonal matrix $\hat{D}$ are
positive. We can then define the inverse square root of
$\hat{{\cal M}}_{II}$ as 
\be
\hat{{\cal M}}_{II}^{-1/2}\equiv \hat{\Ome}_I^T\hat{D}^{-1/2}\hat{\Ome}_I . 
\ee 
The inverse square root of $\hat{{\cal M}}_{OO}$ is defined similarly,
\be
\hat{{\cal M}}_{OO}^{-1/2}\equiv \hat{\Ome}_O^T
\hat{B}^{-1/2}\hat{\Ome}_O 
\ee
where $\hat{{\cal M}}_{OO}
=\hat{\Ome}_O^T\hat{B}\hat{\Ome}_O$, $\hat{\Ome}_O$ is orthogonal,
and $\hat{B}$ is diagonal and positive.

\section{Prove that $\hat{\Lambda}$ is a positive matrix,
  $\lambda\le 1$, and others}

To simplify the proof, we perform an orthogonal transformation
$\hat{\Ome}_O\otimes\hat{\Ome}_I$ on the forces. To keep the currents
conjugated with forces, the same transformation must be exerted on the
currents. The transformation diagonalize the matrix 
$\hat{{\cal M}}_{II}$ and $\hat{{\cal M}}_{OO}$. As both of them are
positive matrix we can further perform the following
transformation
\be
{\cal F}_n \to {\cal F}_n \sqrt{{\cal M}_{nn}} , \quad {\cal J}_n \to
\left. {\cal J}_n \right/ \sqrt{{\cal M}_{nn}} . \label{shrink}
\ee
This leads to
\be
{\cal M}_{nk}\to \frac{{\cal M}_{nk}}{\sqrt{{\cal M}_{nn}{\cal
      M}_{kk}}} .
\ee
After the above transformation the matrix $\hat{{\cal M}}_{II}$ and 
$\hat{{\cal M}}_{OO}$ become identity matrix. Now for the real matrix
$\hat{{\cal M}}_{IO}$ there always exists a decomposition 
$\hat{{\cal M}}_{IO}=\hat{\ome}_I^T\hat{C}\hat{\ome}_O$ where
$\hat{\ome}_I$ and $\hat{\ome}_O$ are orthogonal matrices and
$\hat{C}$ is a diagonal matrix (but no need to be a square matrix)
(see Ref.~\cite{s-matrix}).
Performing the orthogonal transformation $\hat{\ome}_O\otimes
\hat{\ome}_I$ on the forces and currents and using Eq.~(\ref{matrix}),
we obtain
\be
\hat{\Lambda} = \hat{{\cal M}}_{IO}\hat{{\cal M}}_{IO}^T =
\hat{C} \hat{C}^T . \label{lam-ap}
\ee
Now $\hat{\Lambda}$ is a diagonal matrix with all diagonal
elements greater than or equal to zero. We thus proved that the
coupling matrix $\Lambda$ is a positive matrix. The largest eigenvalue
of the coupling matrix $\hat{\Lambda}$ is also positive, i.e.,
$\lambda\ge 0$. Labeling the diagonal elements of $\hat{C}$ as $y_n$
($n=1, \dots N$ is integer if the dimension of the matrix $\hat{C}$ is
$N\times N^\prime$ with, say, $N\ge N^\prime$), the
Onsager matrix now becomes
\be
{\cal M} = \left( \begin{array}{cccccccccc}
    1 &  & 0  & y_1 &  & 0 &  & 0 \\
     & \ddots &  & & \ddots &  &\ddots &  \\
     0 & & 1 & 0 & & y_N &  & 0 \\
     y_1 & & 0 & 1 & & 0 &  & 0 \\
     & \ddots &  & & \ddots & &\ddots &  \\
    0 & & y_N & 0 & & 1 &  & 0 \\
     &\ddots &  &  &\ddots & & \ddots & \\
    0 &  & 0 & 0 &  & 0 & & 1 \\ 
  \end{array}\right) . \label{diag-form}
\ee
It follows from Eqs.~(\ref{lam-ap}) and (\ref{phim2}) that
\be
\lambda = {\rm max}\{y_n^2\} . \label{maxyn}
\ee
According to the second law of thermodynamics all eigenvalues of the
Onsager matrix are positive, i.e.,
\be
1+y_n\ge0, \quad 1-y_n\ge 0, \quad \forall n ,
\ee
according to Eq.~(\ref{diag-form}). Therefore $0\le \lambda\le 1$ and
the figure of merit $\xi=\lambda/(1-\lambda)$ is positive definite.

At this point one can also show that when a machine is operating in a
reverse way, i.e., the output channels become input channels and vice
versa. The matrix $\hat{\Lambda}$ becomes
$\hat{\Lambda}=\hat{C}^T\hat{C}$ which has the same largest eigenvalue
as before. In this way we proved that when a machine is operated in a
reverse way the degree of coupling $\lambda$ and the figure of merit
does not change.

Finally from Eq.~(\ref{diag-form}) one can also directly show that 
$\hat{{\cal M}}_{IO}\hat{{\cal M}}_{OO}^{-1}\hat{{\cal
    M}}_{OI}=\hat{C} \hat{C}^T$ is positive matrix (i.e.,
all its eigenvalues are positive). Therefore the
largest eigenvalue of $\hat{{\cal M}}_{IO}\hat{{\cal M}}_{OO}^{-1}
\hat{{\cal M}}_{OI}$ is positive, i.e., $\Upsilon>0$.

\section{Thermodynamic bounds for systems with asymmetric Onsager matrix}
We shall focus on the situation considered in the main text where
$\hat{{\cal M}}_{OI}=r \hat{{\cal M}}_{IO}^T$. For this situation
one can perform the same transformation as in previous section:
symmetric matrices ${\cal M}_{II}^s$ and ${\cal M}_{OO}^s$ can be
diagonalized by orthogonal transformations; after that performing the
transformation (\ref{shrink}) and another orthogonal transformation
${\cal M}_{II}^s$ and ${\cal M}_{OO}^s$ become identity matrices and
${\cal M}_{IO}\to \hat{C}$, ${\cal M}_{OI}\to r \hat{C}^T$. The
second law of thermodynamics requires that all eigenvalues of
$\hat{{\cal M}}^s$ are greater than or equal to zero. Therefore
\be
1 - \frac{1}{2} (1 + r) y_n \ge 0, \quad  1 + \frac{1}{2} (1 +
r) y_n \ge 0, \quad \forall n .
\ee
The degree of coupling is given by
\be
\lambda = r~ {\rm max}\{y_n^2\} .
\ee
Therefore 
\be
0\le \lambda \frac{(1+r)^2}{4r} \le 1 .
\ee

The discussions in Sec.~\ref{sec:asym} can be generalized to the situation when 
${\cal M}_{OI}$ is not proportional to ${\cal M}_{IO}$ but they can
still be diagonalized simultaneously by an orthogonal transformation.
The diagonal form of $\hat{{\cal M}}_{IO}$ is $diag\{y_n\}$ while
that of $\hat{{\cal M}}_{OI}$ is $diag\{r_ny_n\}$. The optimal
exergy efficiency is given by
\be
\phi_{max} = {\rm
  max}\left\{r_n\frac{\sqrt{\xi_n+1}-1}{\sqrt{\xi_n+1}+1}\right\}
,
\ee
where
\be
\xi_n \equiv \frac{\lambda_n}{1-\lambda_n}, \quad \lambda_n \equiv r_n y_n^2 .
\ee
And the output power at maximum exergy efficiency is
\be
\dot W (\phi_{max}) = 
  \frac{1}{4}( 1 - r_n^{-2} \phi_{max}^2 ) r_n\lambda_n \left(\vec{{\cal
    F}}_I^T\hat{{\cal M}}_{II}\vec{{\cal F}}_{I}\right) 
\ee
for the $n$ that maximizes the efficiency.
The maximum output power is
\be
\dot W_{max} = \frac{1}{4} {\rm max}\left\{  r_n\lambda_n \right\} \left(\vec{{\cal
    F}}_I^T\hat{{\cal M}}_{II}\vec{{\cal F}}_{I}\right) .
\ee
The optimal exergy efficiency for maximum power is given by
\be
\phi_{opt}(\dot W_{max}) = \frac{r_{n^\prime}
  \xi_{n^\prime}}{2(\xi_{n^\prime} + 2)} 
\ee
for the $n^\prime$ that maximizes the output power (which may be
different from that maximizes the efficiency). As $n$ can be different
from $n^\prime$, the relationship between the two optimal efficiencies
and powers can be more complicated then we discussed in the main text.

\end{appendix}


\begin{thebibliography}{99}
\bibitem{note} Note that the reversed logic does not hold. That is, it
  is not true that zero determinant of Onsager matrix implies 100\%
  efficiency. This is because zero determinant of Onsager matrix can
  also represent zero response to certain external forces which
  deliver zero energy conversion.
\end{thebibliography}
\end{document}